\documentclass{iopart}
\usepackage{epsfig}

\newcommand{\bk}{{\bf k}}
\newcommand{\eps}{{\tilde \epsilon}}
\newcommand{\trho}{{\tilde \rho}}
\newcommand{\rS}{{\mathcal S}}
\newcommand{\V}{{\mathcal V}}
\newcommand{\N}{{\mathcal N}}

\newcommand{\Z}{{\mathcal Z}}
\newcommand{\tZ}{{\tilde {\mathcal Z}}}
\newcommand{\tZE}{{\tilde {\mathcal Z}_{\rm ex}}}
\newcommand{\E}{{\mathcal E}}
\newcommand{\cP}{{\mathcal P}}

\begin{document}

\jl{1}
\title[Universal thermodynamics and its consequences]{
Gases in two dimensions: universal thermodynamics and its consequences}

\author{Drago\c s-Victor Anghel \footnote{E-mail: dragos@fys.uio.no}} 

\address{University of Oslo, Department of Physics, P.O.Box 1048 
Blindern, N-0316 Oslo, Norway, {\em and} NIPNE  -- ``HH'', P.O.Box MG-6, 
R.O.-76900 Bucure\c sti - M\u agurele, Romania}

\begin{abstract} 
I discuss ideal and interacting quantum gases obeying general 
fractional exclusion statistics. 
For systems with constant density of single-particle states, 
described in the mean field approximation, the entropy depends neither 
on the microscopic exclusion statistics, nor on the interaction. 
Such systems are called {\em thermodynamically equivalent} and I show that 
the microscopic reason for this equivalence is a one-to-one correspondence 
between the excited states of these systems. This provides a method, 
different from the bosonisation technique, to transform 
between systems of different exclusion statistics. In the last section 
the macroscopic aspects of this method are discussed.

In Appendix A I calculate the fluctuation of the ground 
state population of a condensed Bose gas in grandcanonical ensemble and 
mean field approximation, while in Appendix B I show a situation where 
although the system exhibits fractional exclusion properties on 
microscopic energy intervals, a rigorous calculation of the population of 
single particle states reveals a condensation phenomenon. This also 
implies a malfunction of the usual and simplified calculation technique 
of the most probable statistical distributions.
\end{abstract}
\pacno{05.30.Ch, 05.30.Pr, 05.30.Jp, 05.70.-a}
\submitto{\JPA}
\maketitle

\section{Introduction}
\label{intro}

Considerable interest has been shown in the recent years to the 
study of particle systems that exhibit 
fractional exclusion statistics (FES) -- model introduced by Haldane 
in Ref. \cite{haldane} and which applies, among other, to 
quasiparticle excitations at lowest Landau level in the fractional 
quantum Hall effect and spinon 
excitations in a spin-$\frac{1}{2}$ quantum antiferomagnet \cite{haldane}.
The general thermodynamic properties of these systems 
have been deduced mainly by Isakov \cite{isakov1,isakov3} and Wu \cite{wu}. 
Isakov also showed that anyons \cite{leinaas} and one-dimensional (1D)
systems of particles described by thermodynamic Bethe Ansatz (TBA)
\cite{TBA1,TBA2,sutherland} have the same thermodynamic behaviour as systems 
exhibiting FES. Since a similar thermodynamic behaviour of two 
systems implies a similarity between the {\em excitation} spectra 
and eventually -- depending on the ensemble in which the thermodynamics 
is discussed -- the same ground state energies as functions of the 
particle numbers, this is irrelevant for comparing the microscopic 
properties of the constituent particles. 
%
%Although a similar thermodynamic behaviour of two 
%systems must imply a similarity of the {\em excitation} spectra 
%at least in the thermodynamic limit, 
%this is irrelevant for comparing the microscopic 
%properties of the constituent particles. 
%
Sutherland and Iguchi generalized the concept of Bethe Ansatz from 
1D to two and three dimensions (2,3D), showing that bosons and 
fermions described by this model exhibit also at microscopic level 
fractional exclusion statistics \cite{sutherland1}.

The Bethe Ansatz was first used to calculate hamiltonian eigenvectors 
of 1D spin chains with anisotropic spin-spin interaction (see for example 
Ref. \cite{yang} and references therein) and then applied, starting with 
Lieb and Liniger \cite{TBA1}, to interacting 
trapped particles. The model applies in general to dilute systems, 
with short-range two-body interaction. 
In the other extreme are dense 
systems with long range interaction -- as compared to interparticle 
distance --, which may be described in the 
mean field approximation (MFA). Nevertheless, MFA is applied 
also to dilute systems. 

Murthy and Shankar were the first to modify the MFA model for a Fermi system 
\cite{murthy} -- let me call this new model MFA' -- 
by redistributing the mean field interaction among single particle 
energies, so that a particle of energy $\epsilon$ is assumed to 
interact just with particles of energy $\epsilon' < \epsilon$ 
(see \ref{mess}). 
Within the MFA' model, the interacting Fermi system is equivalent to an 
ideal Haldane system of statistics parameter $\alpha$ which depends 
on the interaction strength. 
Independently, 
Hansson et al. \cite{hansson0} used the MFA' model to define single 
particle energies in a field theory of anyons in the lowest Landau level, 
while Isakov and Viefers \cite{isa-vif}, among other things, showed 
again that this model reproduces Haldane's FES.
All these results have been obtained for constant DOS in the 
ideal systems. 

More recently, Bhaduri et al. \cite{bhaduri} showed that also a 
repulsively interacting 2D MFA Bose gas has identical thermodynamic 
parameters as an ideal FES gas with $\alpha$ 
again fixed by the interaction strength. 
Continuing Ref. \cite{bhaduri}, Hansson et al. \cite{hansson} discussed 
the applicability of the MFA to bosons with repulsive 
delta function interaction, in quasi 2D traps, and 
transformed again the MFA into MFA' to calculate the thermodynamic 
parameters of the corresponding FES gas.
%
%along the same lines as in Refs. 
%\cite{murthy,hansson0,isa-vif}, Hansson et al. \cite{hansson} 
%transformed again the MFA into MFA', calculated the thermodynamic 
%parameters of the corresponding FES gas, and discussed the applicability 
%of the MFA to 

In what follows I shall say that two systems are {\em thermodynamically 
equivalent} if they have the same entropy as a function of temperature, 
at fixed volume (or external potential) and particle number. 
The amazing thermodynamic similarity in 2D between interacting Bose or 
Fermi systems, and ideal FES systems with properly chosen statistics 
parameter $\alpha$ rests on their thermodynamic equivalence. %which is 
%a consequence of the one-to-one correspondence between excited states 
%of systems of any microscopic exclusion statistics (see Section 
%\ref{micro}). 
%

To show the special character of nonrelativistic 2D ideal systems, 
I will start by deriving unified expressions 
for their thermodynamic quantities in terms of polylogarithmic functions 
\cite{lewin}.
This bridges Lee's description of bosons and fermions \cite{lee1,lee2} 
through the intermediate statistics of haldons and emphasizes in 
a most simple way the thermodynamic equivalence of the systems with 
constant DOS of any statistics. 
The thermodynamic equivalence of 2D Bose and Fermi systems was 
proved by several authors \cite{lee3,lee4,apostol,viefers,may}.
The fact that the temperature dependence of the thermodynamic potentials 
of Haldane systems with constant DOS is independent of 
statistics was observed before \cite{isakov2}. Here I merely identify the 
functions involved, which makes the writing and manipulations of 
the thermodynamic quantities much easier and compact.
In Section \ref{micro} I give the microscopic explanation 
of this equivalence, which rests {\em not on a microscopic similarity 
between systems, but on a one-to-one correspondence between the excited states 
of these systems}. This also gives us a transformation method between 
systems of different statistics, which will be extended to more 
general spectra elsewhere \cite{viitor2}. 
In Section \ref{macro}, based only on macroscopic arguments, I show that 
all the thermodynamic quantities describing 2D systems in the 
Thomas-Fermi approximation and in any 
trapping potential, depend only on the temperature and the ground 
state energy density, which is a function of particle density. 
This leads to the definition of classes of thermodynamically 
equivalent systems. Since the proof is based only on macroscopic
arguments, it shows explicitly that similar thermodynamic behaviour 
does not imply microscopic similarity. 
Based on Sections \ref{ideal} and 
\ref{general}, I can say that the transformation from MFA to MFA' 
merely sets an abstract point of view or counting rule. 
In \ref{fluctuations} I calculate the fluctuation of the ground 
state population of a condensed Bose gas in MFA approximation and 
finally, in \ref{mess} I compare in more detail the results given 
by MFA and MFA' models. I show that in the case of interacting bosons 
in 2D boxes, depending on the choice of a parameter, MFA' may lead 
to condensation on the lowest energy level, in contrast 
to MFA. 

\section{Ideal Haldane systems with constant density of states}
\label{ideal}

\subsection{Thermodynamic equivalence}

In what follows I shall consider Haldane systems with DOS of the form 
$\sigma(\epsilon) = C \epsilon^s$ ($s > -1$), where $C$ is a constant 
and $\epsilon$ is the single particle energy. If the system 
(of nonrelativistic particles) is in a $d$-dimensional ($d$D) container 
with no external fields, $C$ is proportional to the hypervolume and 
$s=d/2-1$. 
The exclusion statistics, characterized by the energy independent 
parameter $\alpha\ge 0$, is manifested between particles within the same 
infinitesimal energy interval $\delta\epsilon$ \cite{wu}. 
The values $\alpha=0$ and $\alpha=1$ correspond to 
Bose and Fermi statistics, respectively.
With these notations, the average population of a single particle 
state is \cite{wu}
\begin{equation} \label{n}
n(\epsilon) = \{w(\zeta_\epsilon) + \alpha \}^{-1} , 
\end{equation}
where $w$ satisfies the equation 
$w(\zeta_\epsilon)^\alpha[1+w(\zeta_\epsilon)]^{1-\alpha}=\zeta_\epsilon^{-1} \equiv \rme^{(\epsilon - \mu)/k_{\rm B}T}$, in obvious notations (see 
\ref{mess} for a special case). 
The grand canonical thermodynamic potential, 
$\Omega \equiv -PV = k_{\rm B}T \int_0^\infty \rmd\epsilon\, C\epsilon^s \log{\{(1-\alpha n)/[1+(1-\alpha)n]\}}$ 
\cite{wu}, can be put in the form
\begin{equation}\label{pv}
PV = \frac{1}{s+1}\int_0^\infty \rmd\epsilon\, C\epsilon^{s+1} n(\epsilon) 
\equiv \frac{U}{s+1} \, ,
\end{equation}
where $U$ is the internal energy. The total number of particles is 
$N = \int_0^\infty \rmd\epsilon\, C\epsilon^s n(\epsilon)$. 
All these functions may be calculated by expressing $\epsilon$ in 
terms of $w$, %: 
%$\epsilon = k_{\rm B} T \log{[w^\alpha (1+w)^{1-\alpha}]} + \mu$, 
but the integrals cannot be performed analytically for general $s$ and $T$. 
%Isakov et al. \cite{isakov2} gave low temperature and low density 
%expansions of the thermodynamic functions. 
Nevertheless, for $s=0$ ($\sigma\equiv C$), all the thermodynamic 
quantities can be expressed in terms of elementary or polylogarithmic 
functions. I start with 
\begin{equation} \label{N}
N = k_{\rm B}T \sigma \log{(1+y_0)} ,
\end{equation}
where $y_0$ satisfies the equation 
$(1+y_0)^{1-\alpha}/y_0 = \zeta^{-1} \equiv \rme^{-\mu/k_{\rm B}T}$.
%We denote  $w_0 \equiv 1/y_0$. 
Note that $y_0$ depends on $N$ and $T$, but not on $\alpha$. 
Moreover, $0=y_0(N,T=\infty)\le y_0(N,T)\le y_0(N,T=0)=\infty$. 
From Eq. (\ref{N}) follows
\begin{equation}\label{mu}
\exp{[(\mu - \alpha N/\sigma)/k_{\rm B} T]} = 1- \exp{[-N/(\sigma 
k_{\rm B}T)]}. 
\end{equation}
where I identify the (generalized) Fermi energy as 
$\epsilon_{\rm F} \equiv \lim_{T\to 0} \mu = \alpha N/\sigma$ and observe 
that $\mu - \epsilon_{\rm F}$ is also independent of $\alpha$.
After some algebra I get the desired expression for $\Omega$ and $U$:
\begin{equation} \label{omega1}
\Omega = -U = (k_{\rm B}T)^2 \sigma\left[\frac{1-\alpha}{2} 
\log^2{(1+y_0)} + Li_2(-y_0)\right] ,
\end{equation}
where $Li_2$ is Euler's dilogarithm \cite{lewin}. Using the relation 
$Li_2(x)+Li_2[-x/(1-x)] = -(1/2)\log^2{(1-x)}$, valid for any 
$x<1$ \cite{lewin}, one can prove that $\Omega \le 0$ for any 
$\alpha \ge 0$ and $y_0 \ge 0$, as expected. Combining Eqs. 
(\ref{N}) and (\ref{omega1}), it follows:
\begin{equation}\label{echiv} 
\Omega = -U = \frac{1-\alpha}{2}\frac{N^2}{\sigma} + 
(k_{\rm B} T)^2 \sigma Li_2(-y_0) ,
\end{equation} 
In Eqs. (\ref{N}) and (\ref{echiv}) the equivalence between ideal gases 
obeying {\em any} statistics is highlighted in the simplest way. Since 
$y_0$ does not depend on statistics, but just on $N$ and $T$, it is 
obvious that the difference between the thermodynamic potentials 
of particles with different $\alpha$s comes just from an additive 
constant. All the temperature dependence is the same. 
As an example, one can set $\alpha = 0$ and $\alpha = 1$ and use 
the Landen's relations \cite{lewin} to re-obtain the familiar Bose 
and Fermi thermodynamic potentials \cite{lee3,lee4}. 
Making use of Eqs. (\ref{echiv}) and (\ref{N}), one can obtain in the 
usual way expressions for the 
entropy and specific heat, which are both independent of $\alpha$:
\begin{eqnarray}
S &=& -k_{\rm B}^2 T \sigma [2Li_2(-y_0) + \log{(1+y_0)}\log{y_0}] 
\label{entropia} \\
%&=& -k_{\rm B}^2 T C [Li_2(-y_0) - Li_2(1+y_0) - i\pi \log{(1+y_0)} + 
%\frac{\pi^2}{6}] \nonumber \\
C_{\rm V} &=& - \frac{N^2}{T\sigma}\frac{1+y_0}{y_0}- 2k_{\rm B}^2 
T\sigma Li_2(-y_0) . \label{cv} 
\end{eqnarray}
Since according to Eq. (\ref{N}) $y_0\to \infty$ when 
$T\to 0$, making use of the asymptotic behaviour of the dilogarithm, 
$Li_2(-y_0) \sim -[\pi^2/6 + \log^2(y_0)/2]$ one can show that 
in the limit of low temperatures 
$C_{\rm V} \sim (\pi^2/3)k_{\rm B}^2T\sigma\{1- O[\log^2(1+y_0)/y_0]\}$. 
In the case of particles inside 2D boxes, 
the leading term is identical to the result for fermions obtained by 
Li et al. \cite{li}.

\subsection{Thermodynamic equivalence from the microscopic point of view: 
Haldane--Bose transformation}
\label{micro}

\begin{figure}[t]
\begin{center}
\unitlength1mm\begin{picture}(35,50)(0,0)
\put(0,0){\epsfig{file=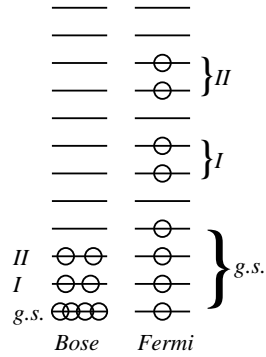,width=35mm}}
\end{picture}
\caption{Excitation in Bose and Fermi gases with the same equidistant 
single-particle energy levels. The fermions in the g.s., I, II, \ldots  
groups are in correspondence with the bosons on the ground state, 
first, second, \ldots excited states, respectively. In this way I
establish a one-to-one correspondence between the particle configurations 
in the Bose and Fermi systems, with the same excitation energies.}
\label{spectru}
\end{center}
\end{figure}

Now let me inspect this equivalence from the microscopic point of view.
%The equivalence just proved holds also at microscopic scale. 
For this I consider a Haldane and a Bose system with the same 
particle number, $N$, and DOS, $\sigma\equiv C$. 
%We further assume that in the first system 
%the exclusion statistics is manifested between 
%particles within the same infinitesimal energy interval $\delta\epsilon$, 
%with $\alpha$ independent of energy \cite{wu}. 
In the Haldane system I divide the energy axis into intervals of equal 
length, $\delta\epsilon$, 
and number them, starting from zero, at the lowest interval. 
Each of these intervals contains the same number of single particle states, 
$d=\sigma\delta\epsilon$, and a variable number of particles, 
$n_{{\rm H},i}$ ($\ i\ge 0$). For state counting 
purposes I define a ``Bose dimension'' of the subspace corresponding 
to the interval $i$ as $d_{{\rm B},i}=d-\alpha (n_{{\rm H},i}-1)$, so the 
total size of its Hilbert space is 
$w_{{\rm H},i}=(d_{{\rm B},i}+n_{{\rm H},i}-1)!/[n_{{\rm H},i}!(d_{{\rm B},i}-1)!]$ \cite{haldane}. 
I denote $\bar N_{{\rm H},i}=\sum_{i=0}^{i-1}n_{{\rm H},j}$ and 
${\mathcal E}_{{\rm H},i}=n_{{\rm H},i}(i\cdot\delta\epsilon-\bar \epsilon_{{\rm F},i})$, 
where ${\mathcal E}_{{\rm H},i}$ shall be called the excitation energy of 
the particle group $n_{{\rm H},i}$, while 
$\bar \epsilon_{{\rm F},i}=\alpha\bar N_{{\rm H},i}/\sigma$ is the 
(generalized) Fermi energy of a similar system, but consisting only 
of $\bar N_{{\rm H},i}$ particles 
(I disregarded the distribution of particles inside the $i^{\rm th}$ 
energy interval). 
Obviously, the ground state energy of the Haldane system is 
$U_{{\rm H},0}=\sum_{i=0}^\infty n_{{\rm H},i}\bar\epsilon_{{\rm F},i}$ 
while the total energy is 
$U_{\rm H}=\sum_{i=0}^\infty n_{{\rm H},i}\epsilon_{{\rm H},i}=\sum_{i=0}^\infty {\mathcal E}_{{\rm H},i}+U_{{\rm H},0}$. 
Under canonical conditions, the weight of such a distribution is 
${\mathcal W}_{{\rm H},\{n_i\}}=[\prod_i\exp{(-{\mathcal E}_{{\rm H},i}/k_{\rm B}T)}w_{{\rm H},i}]/Z_{{\rm H},N,T}$, 
where $Z_{{\rm H},N,T}$ is the partition function and satisfy the condition 
$\sum_{\{n_i\}}{\mathcal W}_{{\rm H},\{n_i\}}=1$; the sum is taken over 
all distributions $\{n_i\}$.
Now let me put the distributions in the Haldane system in correspondence 
with distributions in the Bose system and calculate the new weights. 
For this, given a distribution $\{n_i\}$ in the Haldane system, I 
divide the energy axis of the Bose system in 
nonequal intervals, $\delta\epsilon_{{\rm B},i}=d_{{\rm B},i}/\sigma$, 
and place $n_i$ bosons in each of them. As a consequence, the 
size of the Hilbert space of interval $i$ is $w_{{\rm B},i}=w_{{\rm H},i}$. 
Since in a degenerate Haldane gas, in an energy interval $1/\sigma$ 
coexist on average $1/\alpha$ particles, to obtain the correct energy 
distribution one has to overlap $\alpha/\sigma$ of any consecutive 
energy intervals above. 
Then, the excitation energy of the group $n_i$ is 
$n_i\sum_0^{i-1}\delta\epsilon_{{\rm B},j}=n_i\sum_0^{i-1}[(d-\alpha n_j)/\sigma]\equiv {\mathcal E}_{{\rm H},i}$. Therefore, the two configurations have 
identical statistical weights. To prove that the two systems are 
equivalent, I have to show 
that also, to any configuration in the Bose system it corresponds one 
configuration in the Haldane system, with the same excitation energy and 
statistical weight. But this can be done just following the steps outlined 
above in reverse order. 
In conclusion I showed that there is a one-to-one correspondence between 
the configurations in the Bose and Haldane systems. These configurations 
have the same ``excitation energy'' and the same statistical weights, 
so the two systems are statistically equivalent even at microscopic level.
This analogy finds its simplest illustration in the case of 
ideal Bose and a Fermi gasses with the same spectrum, which consist 
of nondegenerate, equidistant single particle energy levels (1D harmonic 
trap), as shown in Fig. \ref{spectru}.

From the equivalence proven above and the fact that a macroscopic 
Bose system with $\sigma\equiv C$ does not experience Bose condensation 
(see Ref. \cite{holthaus} for interpretations), I conclude that 
at any finite temperature the Fermi sea is not still (excitations occur 
at any depth in the Fermi sea, in macroscopic number). On the other hand, 
by transforming a Bose into a Fermi system, one may describe it 
(and eventually calculate interaction effects) just by considering 
low energy particle-hole excitations around the Fermi surface (bosonisation), 
which may considerably simplify the calculations. 

\section{Generalization: interacting systems in arbitrary traps}
\label{general}

\subsection{Statistical mechanics in boxes}\label{boxes}

I investigate here the effects of particle-particle interaction in 
systems inside $d$-dimensional ($d$D) containers with no external fields. 
Next subsection will consider arbitrary systems.
%I use Landen's relations \cite{lewin} to write Eq. (\ref{omega1}), 
%corresponding to ideal haldons, in the form 
%$U^{(0)}_\alpha\equiv U^{(0)}_{\alpha,{\rm g.s.}}+\Omega_{\rm B}$, 
%where $U^{(0)}_{\alpha,{\rm g.s.}} = \alpha N^2/2\sigma$ is the 
%``ideal'' ground state energy, while 
%\begin{eqnarray} 
%$\Omega_{\rm B}\equiv-U_{\rm B}=-(k_{\rm B}T)^2\sigma Li_2[y_0/(1+y_0)]$
%\end{eqnarray}
%is the ``Bose'' thermodynamic potential. 
%[in Eq. (\ref{omegaB}) we made use of Landen's relations \cite{lewin}]. 
In the Thomas-Fermi mean field approximation (TF-MFA), the 
particle-particle interaction is replaced by an 
effective one-body potential, $H_{\rm I}(N)$ (I assume the 
interaction does not lead to a phase separation) and the total energy of 
the system is  
%
%\begin{eqnarray}
\begin{equation} \label{conf_en} 
E_{\alpha,\{\bk\}}=\sum_{\{\bk\}}\epsilon_\bk + 
\frac{NH_{\rm I}(N\sigma^{-1})}{2} 
\equiv E_{\rm ex}+U^{(0)}_{\alpha,{\rm g.s.}}+ 
\frac{NH_{\rm I}(N)}{2} ,
\end{equation}
%\end{eqnarray}
where $\epsilon_\bk$ are the single particle energies %of the ideal system 
and $U^{(0)}_{\alpha,{\rm g.s.}}$ is the ground-state energy of the 
system without interaction. 
The quantum numbers that specify the single particle states are denoted 
by $\bk$ (in the case of spinless particles, $\bk$ is the momentum). 
If I denote by $E^{(0)}_{\alpha,\{\bk\}}\equiv\sum_{\{\bk\}}\epsilon_\bk$ 
the energy of the ideal system, then the excitation energy 
(for 2D boxes this is the energy of the equivalent Bose system, as 
defined in Section \ref{micro}) is 
$E_{{\rm ex},\{\bk\}}\equiv E^{(0)}_{\alpha,\{\bk\}}-U^{(0)}_{\alpha,{\rm g.s.}}$. 
In the microcanonical ensemble $N$ is fixed, so the partition 
function is just a function of the excitation energy $E_{\rm ex}$: 
$\bar Z_{\alpha,N}(E_{\rm ex})$. 
The canonical partition function is usually written as 
\begin{eqnarray}
Z_{\alpha,N}(\beta) &=& \int_{U_{\alpha,{\rm g.s.}}}^\infty \rmd E_\alpha
\rme^{-\beta E_\alpha}\bar Z_{\alpha,N}(E_{\rm ex}) \label{Z_int} \\
&=&\rme^{-\beta NH_{\rm I}(N)/2}
\underbrace{\int_{U^{(0)}_{\alpha,{\rm g.s.}}}^\infty 
\rmd E^{(0)}_\alpha \rme^{(-\beta E^{(0)}_\alpha)}
\bar Z_{\alpha,N}(E_{\rm ex})}_{Z^{(0)}_{\alpha,N}(\beta)} \nonumber \\
%\label{Z_0a} \\
&=&\rme^{-\beta U_{\alpha,{\rm g.s.}}} \underbrace{\int_0^\infty 
\rmd E_{\rm ex} \rme^{-\beta E_{\rm ex}}
\bar Z_{\alpha,N}(E_{\rm ex})}_{Z^{(0)}_{0,N}(\beta)} \nonumber 
%\label{Z_00}
\end{eqnarray}
where I used the shorter notations $\beta\equiv(k_{\rm B}T)^{-1}$ and 
$U_{\alpha,{\rm g.s.}}\equiv U^{(0)}_{\alpha,{\rm g.s.}}+\frac{1}{2}NH_{\rm I}(N)$. 
I also denoted by $Z^{(0)}_{\alpha,N}(\beta)$ (or $Z^{(0)}_{\alpha,N}(T)$) 
the canonical partition function 
of the ideal Haldane gas of parameter $\alpha$. $Z^{(0)}_{0,N}(\beta)$ 
corresponds to an equivalent Bose gas of excitations. The coefficients in 
front of $Z^{(0)}_{\alpha,N}(\beta)$ and $Z^{(0)}_{0,N}(\beta)$ in the two 
expressions for $Z_{\alpha,N}(\beta)$ are redundant in the 
canonical ensemble, but they should not be ignored 
in grandcanonical calculations.

Let me now find the population of the single particle energy levels. 
Since by changing the single particle quantum numbers $\bk$ into $\bk'$ 
the total energy of the system changes by $\epsilon_\bk'-\epsilon_\bk$, 
the relative occupation probability of the states $\bk$ and $\bk'$ 
is the same as in the case of the ideal Haldane system of parameter $\alpha$. 
These occupation probabilities in the canonical ensemble
follow from $Z^{(0)}_{\alpha,N}(\beta)$ which may be calculated by the 
usual saddle point method from the grandcanonical ensemble of the ideal 
system (after dropping the term $NH_{\rm I}(N)/2$ from the expression 
of total energy) -- see Ref. \cite{holthaus0} for 
the special case of Bose-Einstein condensed systems. 
In this way I obtain the average population of the state $\bk$, 
which is given by Eq. 
(\ref{n}) with $\epsilon\equiv\epsilon_\bk$ and the {\em apparent} 
chemical potential, $\mu_{\rm a}$, which in 2D satisfies Eq. (\ref{N}). 
Note also that $\mu_{\rm a}$ is not the {\em real} chemical 
potential, which is defined as
\begin{equation} \label{mur}
\mu \equiv -k_{\rm B}T 
\frac{\partial \log (Z_{\alpha,N}(T))}{\partial N} 
= \frac{\rmd U_{\alpha,{\rm g.s.}}}{\rmd N}
\underbrace{-k_{\rm B}T 
\frac{\partial \log (Z^{(0)}_{0,N}(T))}{\partial N}}_{\equiv\mu_{\rm B}} .
\end{equation}
Taking out the ground state energy from the total energy of the system, 
as done in the expressions (\ref{Z_int}) exposes the contribution 
to the partition function coming just from the excitations, which 
have a bosonic character. Then $\mu_{\rm B}$ 
may be called the ``chemical potential'' of these excitations. 
Although $\lim_{T\to0}Z^{(0)}_{0,N}(T)=1$, for 
any $N$ (I assumed nondegenerate ground state), 
implies $\lim_{T\to0}\mu_{\rm B}(T)=0$, unlike the 
chemical potential of an ideal Bose system, $\mu_{\rm B}$ may take also 
positive values, as for example in the case of ideal 1D fermions. 
In the case of particles in a 2D box, 
the excitation partition function is just the partition function of the 
equivalent Bose system. In \ref{fluctuations} I apply Eqs. (\ref{Z_int}) 
to calculate the fluctuation of the ground state population 
in a nonideal Bose-Einsten condensed system in grandcanonical ensemble, 
while in \ref{mess} I discuss in more detail the MFA' model.

\subsection{General trapping potential: thermodynamic considerations}
\label{macro}

To apply the procedure introduced in Section \ref{boxes} 
%of handling separately the ground state energy and excitation energy of a system, 
from a macroscopic point of view, I shall first show some general properties 
of the entropy function of a neutral fluid, characterized by the 
extensive parameters $U$ (internal energy), $V$ (volume), and $N$ 
(particle number): $S(U,V,N)$. Non-essential 
parts of the proof will be skipped. In what follows, the 
parameters omitted in expressions are the parameters hold fix.
$S$ is a positive, homogenous function of order one, concave 
downwards \cite{pr2} [$S(\lambda U,\lambda V,\lambda N)=\lambda S(U,V,N)$ 
and $S(U_1+U_2,V_1+V_2,N_1+N_2)\ge S(U_1,V_1,N_1)+S(U_2,V_2,N_2)$].
The inverse function, $U(S,V,N)$ is concave upwards. If I write again the 
$U(S,V,N)=U_{\rm g.s.}(V,N)+U_{\rm B}(S,V,N)$, then 
$U_{\rm g.s.}(V,N)\equiv U(S=0,V,N)$ is also concave upwards 
(but, in the general case, $U_{\rm B}$ may not have this property). 
I take $U_{\rm g.s.}(V,N=0)=0$. 
Now I fix $V$. I assume that $N$ and $U(S,N_{\rm fixed})$ take values in the 
intervals $[0,\infty)$ and $[U_{\rm g.s.}(N),\infty)$, respectively 
(the limits may be restricted further, like for example in systems of 
spins or hard-core particles, but here I work with these intervals). 
The concavity property and the range of $U$ implies that $S(U)$ is 
monotonic (increasing). 
If I fix $U\equiv U_{\rm fixed}$ and $U_{\rm g.s.}\equiv 0$, then 
$S(U_{\rm fixed},N)$ is also monotonic in $N$, otherwise is zero at 
both ends of the allowed interval along the $N$ axis. 
If I introduce the function $S_{\rm B}$ by 
$S_{\rm B}(U_{\rm B},V,N)= S(U,V,N)$, then 
$S_{\rm B}$ and $S$ offer two equivalent descriptions of the system. 
Moreover, if $S_{\rm B}$ is concave downwards and coincide with the 
entropy of a new system, say B, then we may say that our 
original system, A, is {\em thermodynamically equivalent} with B 
(I denote it by A$\sim$ B). 
In such a case, if $U_{\rm Cg.s.}(V,N)$ is any homogenous 
function of order one, concave upwards, then the entropy 
$S_{\rm C}(U_{\rm Cg.s.}+U_{\rm B},V,N)= S_{\rm B}(U_{\rm B},V,N)$ 
is a legitimate entropy (with all required properties).
Moreover, if $S_{\rm C}(U_{\rm Cg.s.}+U_{\rm B},V,N)$ describes 
a system C, then A$\sim$B$\sim$C. 
Therefore for any system described by $S(U,V,N)$ I can define the 
{\em Bose entropy} $S_{\rm B}$ by the procedure outlined above. 
According to the definition given in Section \ref{intro}, the systems 
with the same 
$S_{\rm B}$ are called {\em thermodynamically equivalent} and they form 
an equivalence class. 

Since the number of microstates available in a 
system is dependent on the excitation energy $E_{\rm ex}$, $V$, and 
$N$ [$U_{\rm g.s.}(V,N)$ is a redundant quantity], 
one can calculate partition functions 
having as starting point the Bose entropy of the reservoir, 
$\rS_{\rm B}$. In the grandcanonical ensemble, the probability 
associated to a microstate $w(E_{\rm ex},V,N)$ of the system 
is proportional to $\exp[k_{\rm B}^{-1}\rS_{\rm B}(\E_{\rm ex},\V,\N)]$ 
(I use calligraphic letters for the reservoir quantities).
%where the quantities $\E_{\rm ex}$, $\V$, and $\N$ refer to the reservoir. 
Writing as usual the Taylor expansion of the reservoir entropy and 
identifying $U_{\rm B}\equiv E_{\rm ex}$, I obtain the probability of the 
microstate $w$:
\begin{equation}
p(w)=\frac{\rme^{-\beta(U_{\rm B}- \mu_{\rm B}N)}}{\Z} ,
\end{equation}
where $\Z$ is the partition function, 
while $\mu_{\rm B}$ is the chemical potential of the bosonic 
excitations, defined by Eq. (\ref{mur}).

Obviously, the 2D Haldane gases described in MFA form an equivalence 
class. Their Bose entropy is the entropy of the equivalent ideal Bose 
gas. Since $\mu_{\rm B}$ is the same for all systems in an equivalence 
class, the difference between the chemical potentials of different 
systems, is due only to the ground state energy (Eq. \ref{mur}). 
Whether the statistics, the interaction, or both, are responsible 
for the dependence of the ground state energy on $N$, is of no importance. 
Two equivalent gases of different statistics (I showed also in Section 
\ref{boxes} that MFA does not change the microscopic exclusion 
statistics properties), but with the same 
$U_{\rm g.s.}(N,V)$, have identical behaviour even in general trapping 
potentials, as long as the Thomas-Fermi approximation holds (so that 
a system in a variable trapping potential can be regarded as a collection 
of boxes with imaginary walls, in contact with each other).

\section{Conclusions}

To summarize, in the beginning of the paper I gave simple expressions, 
in terms of polylogarithmic functions, 
for the thermodynamic quantities characterizing Haldane ideal systems 
with constant density of single-particle states. These expressions 
are easy to handle and show in a most simple way the thermodynamic 
equivalence (the same entropy function) of such systems. 
Second, I showed the microscopic reason 
for this equivalence, which is the similarity between the excitation 
spectra. 
%of Haldane systems of any exclusion statistics parameter $\alpha$. 
The model used for this purpose is different from the usual 
bosonisation technique (see Ref. 
\cite{marston} for a review) and enables one to transform any Haldane 
system into a Bose system, and vice versa. These transformations: 
Bose$\leftrightarrow$Haldane$\leftrightarrow$Fermi, might be very 
useful if, for a given interaction, techniques developed for one kind of 
systems (like bosonisation for Fermi systems) can be transferred 
to other types of systems (at low temperatures, even in Bose systems, 
the effective thermodynamic contribution comes from a small subset of 
single particle states).

In the Section \ref{general} I gave a general interpretation of this 
correspondence technique, by subtracting from the internal energy of 
the system, $U$, the ground state energy, $U_{\rm g.s.}$. This led to 
an equivalent writing of the partition function, in which the 
internal energy is replaced by the excitation energy, $U-U_{\rm g.s.}$. 
In this formulation the chemical potential 
should be redefined (Eq. \ref{mur}).
Moreover, this redefinition of entropy provides a 
more clear definition of the thermodynamic equivalence, which 
splits the set of physical systems into equivalence classes. 
Systems belonging to the same class have similar excitation spectra 
(at least up to corrections that vanish in the thermodynamic limit). 
In \ref{fluctuations} I apply the results of Section \ref{general} 
to calculate particle fluctuations in a non-ideal, condensed Bose 
system.

In \ref{mess} I show some interesting characteristics of what I called 
the MFA' model and discuss similarities and differences with the 
mean field approximation and fractional exclusion statistics. 
Although on microscopic intervals along the pseudoenergy axis, 
MFA' model reproduces the characteristics of FES gases, as 
defined by Haldane \cite{haldane} and Wu \cite{wu}, in some 
cases a rigorous calculation of the population of 
single particle states reveals a condensation phenomenon which 
implies also a malfunction of the usual calculation technique. 

\ack

I thank Profs. Y. Galperin, J. M. Leinaas, S. Viefers, and M. H. Lee 
for various and fruitful discussions.

\appendix
\section{Particle fluctuation in interacting Bose-Einstein gases} 
\label{fluctuations}

I apply here the technique described in Section \ref{general} to 
calculate the fluctuation of the ground state occupation number, 
$N_0$, of a condensed nonideal Bose gas, in the grandcanonical ensemble.
The interaction is assumed to produce a particle dependent ground 
state energy.
Let me denote by $\Z(\beta,\beta\mu)$ the grandcanonical partition 
function. Using the notations and arguments from Section 
\ref{general} I write 
\[
\fl
\Z(\beta,\beta\mu) = \sum_N \rme^{-\beta (U_{\rm g.s.}-\mu N)} 
\underbrace{\int_0^\infty \rmd E_{\rm ex} \rme^{-\beta E_{\rm ex}}
\bar Z_{N}(E_{\rm ex})}_{Z^{(0)}_{N}(\beta)} \equiv 
\sum_N \Z_N(\beta,\beta\mu) 
\]
and I shall omit the subscripts denoting $\alpha=0$.
$Z^{(0)}_{N}(\beta)$ is the canonical partition function of the 
ideal Bose system. 
As shown in Section \ref{macro}, $U_{\rm g.s.}(N)$ 
is a function concave upwards. 
The maximum of the probability distribution over the 
particle number is given by the equation 
$\partial\Z_N(\beta,\beta\mu)/\partial N=0$, where 
\[
%\begin{equation}
\fl
\frac{\partial\Z_N(\beta,\beta\mu)}{\partial N} = 
\Z_N(\beta,\beta\mu)\left[-\beta\frac{\rmd U_{\rm g.s.}}{\rmd N} + 
\beta\mu - \beta\mu_{\rm B}\right]\equiv 
\Z_N(\beta,\beta\mu)\cdot \Psi_N(\beta,\beta\mu) .
%\end{equation}
\]
The solution of the equation will be denoted by $N_{\rm max}$. 
If there is no $N_{\rm bound}<\infty$, so that 
$\lim_{N\to N_{\rm bound}}\left[\rmd U_{\rm g.s.}/\rmd N\right]=\infty$, 
$N_{\rm max}$ increases to infinity as $\mu$ increases. If the system 
is not pathologic, then $\lim_{N\to\infty}\mu_{\rm B}(N)=0$ and for 
large enough $\mu$, $N_{\rm max}$ is the solution of 
$\rmd U_{\rm g.s.}/\rmd N = \mu$. In such a situation we say that 
the system is Bose-Einstein condensed, since for the relevant 
values of $E_{\rm ex}$ 
the configuration space does not increase with $N$ (the ground 
state is already a particle reservoir).  
The probability distribution $\Z_N$ may be expanded around 
$N=N_{\rm max}$ to get 
$Z_N(\beta,\beta\mu)\approx Z_{N_{\rm max}}(\beta,\beta\mu)\cdot\exp\left[(\partial\Psi_N/\partial N)_{N=N_{\rm max}}(N-N_{\rm max})^2/2\right]$
where 
\[
%\begin{equation}
%\fl
\left.\frac{\partial\Psi_N(\beta,\beta\mu)}{\partial N}\right
|_{N=N_{\rm max}} = 
%\Z_N(\beta,\beta\mu)
%\left\{\left[-\beta\frac{\rmd U_{\rm g.s.}}{\rmd N} + \beta\mu\right]^2 
-\beta\frac{\rmd^2 U_{\rm g.s.}}{\rmd N^2} . %\right\} .
%\end{equation}
\]
For fixed $N$, the fluctuation of $N_0$, $\langle \delta^2 N_0 \rangle_N$, 
is equal to the fluctuation of the particle number on the excited 
energy levels, $\langle\delta^2N_{\rm ex}\rangle_N$ 
(see for example Refs. \cite{holthaus1,holthaus2,navez} for fluctuations 
in ideal gases). 
If the probability distribution of $N_0$ in the canonical ensemble is 
$P(N_0,N)=(2\pi\langle \delta^2 N_0 \rangle_N)^{-1/2}\cdot\exp[-(N_0-\langle N_0\rangle_N)^2/2\langle \delta^2 N_0 \rangle_N]$, 
then the grandcanonical probability distribution of $N_0$ is 
\begin{equation} \label{probability}
\fl
\cP(\beta,\beta\mu,N_0)=\sqrt{\frac{\beta}{2\pi}
\frac{\rmd^2 U_{\rm g.s.}}{\rmd N^2}}\cdot
P(N_0,N)\cdot\exp\left[-\frac{\beta}{2}
\frac{\rmd^2 U_{\rm g.s.}}{\rmd N^2}(N-N_{\rm max})^2\right] .
\end{equation}
Since the system is condensed, the {\em grandcanonical} fluctuation 
of the total particle number, $\langle\delta^2N\rangle$ is equal to 
the fluctuation of the {\em canonical average} ground state occupation 
number. Using this observation and Eq. (\ref{probability}), 
the total fluctuation of $N_0$ in the grandcanonical ensemble 
follows: 
\begin{eqnarray}
\langle\delta^2N_0\rangle_{\beta,\beta\mu}&=& 
\int\rmd N\int\rmd N_0 (N_0-\langle N_0\rangle_{N_{\rm max}})^2
\cP(\beta,\beta\mu,N_0) 
\nonumber \\
&=& \langle\delta^2N_0\rangle_{N_{\rm max}}+\langle\delta^2N
\rangle_{\beta,\beta\mu}
\nonumber
\end{eqnarray}
Therefore the grandcanonic total particle fluctuation of the nonideal 
gas adds quadratically 
to the canonical fluctuation of the ground state population, to give 
the total mean square fluctuation of the number of particle in the 
condensate.

\section{The MFA' model versus mean field approximation and fractional 
exclusion statistics} 
\label{mess}

If the interaction hamiltonian of Section \ref{general} is 
$H_{\rm I}(N)\equiv\sigma^{-1} gN$, 
where $g$ is a dimensionless constant and $\sigma\equiv C$ (constant), 
then (mostly) repeating the arguments from Refs. 
\cite{murthy,hansson0,isa-vif,hansson} I define the MFA' model 
by assimilating the interaction energy from Eq. (\ref{conf_en}) into 
the single particle (kinetic) energies $\epsilon_\bk$, and I define the 
{\em pseudo}energies 
\begin{eqnarray} \label{pseudo}
\eps(\epsilon_\bk)\equiv\epsilon_\bk + \sum_{\bk'} \frac{g}{\sigma}
h(\eps(\epsilon_\bk)-\eps(\epsilon_{\bk'})) n(\bk') ,
\end{eqnarray}
where $n(\bk')$ is the population of the state $\bk'$ and 
$h(x)=1$, $h_0$, or $0$, depending weather $x>0$, $x=0$, or $x<0$, 
respectively; in what follows I assume $g\ge -\alpha$ and $h_0\in [0,1]$, 
where $\alpha$ is the statistics parameter of 
gas without interaction.$^1$\footnotetext{$^1$In the counting rule of 
Hansson et al. \cite{hansson0}, two particles could not have the same 
pseudoenergy even if they occupy the same single particle state. Since I work with a 
(quasi)continuous spectrum, I adopt the more general point of view of 
Isakov and Viefers \cite{isa-vif}, but relaxing also the 
constraint $h_0\equiv\frac{1}{2}$.} 
As Wu \cite{wu}, I divide the energy and pseudoenergy axes into 
infinitesimal (microscopic) intervals. If exist such microscopic 
divisions for which none of the intervals is completely blocked, 
%nor is macroscopically populated, 
then the energy interval 
$[\epsilon_\bk,\epsilon_\bk+\rmd\epsilon)$ is transformed into 
the pseudoenergy interval $[\eps_\bk,\eps_\bk+\rmd\eps)$, where 
$\rmd\eps=\rmd\epsilon+\sigma^{-1}g\rho(\epsilon_\bk)\rmd\epsilon=\rmd\epsilon+\sigma^{-1}g\trho(\eps)\rmd\eps$, and 
$\sigma n(\epsilon_\bk)\rmd\epsilon\equiv\rho(\epsilon_\bk)\rmd\epsilon=\trho(\eps)\rmd\eps$ 
represent the average number of particles in the intervals under 
consideration. Moreover, the Bose dimensions of the two intervals 
$\rmd\eps$ and $\rmd\epsilon$ are: 
\begin{equation} \label{dim}
d_{{\rm B},\rmd\epsilon}=\sigma\rmd\epsilon-\alpha\rho(\epsilon_\bk)
\rmd\epsilon = 
\sigma\rmd\eps-(\alpha+g)\trho(\eps)\rmd\eps \equiv d_{{\rm B},\rmd\eps} ,
\end{equation}
where $\sigma-g\trho(\eps)$ is the {\em effective} density 
of states along the pseudoenergy axis. The Bose density 
$d_{{\rm B},\rmd\eps}$ is identical to the Bose density of an ideal 
Haldane gas of parameter $\alpha+g$ (see also Ref. \cite{murthy2} for 
related discussions). This implies, in accordance with 
Refs. \cite{murthy,isa-vif}, that 
%{\em as long as there is no macroscopically occupied single particle state}, 
the MFA' model leads to the same microscopic particle distribution 
$\trho(\eps)$ as an ideal Haldane gas of statistics parameter $\alpha+g$, 
call it
$n_{\alpha+g}(\eps)$ (\ref{n}) (I included the subscript $\alpha+g$ in the 
notation to specify the statistics parameter of an ideal Haldane gas).
%
%Although the MFA' model is a simple way to describe a system with constant DOS and of total energy of MFA type (\ref{conf_en}) in the second quantization and enables one to do grandcanonical calculations, the pseudoenergies (\ref{pseudo}) have no physical relevance and the {\em real} population of the single particle state $\bk$ of pseudoenergy $\eps(\epsilon_\bk)$ is not given by Eq. (\ref{n}) with $\alpha$ replaced by $\alpha+g$, as it was shown in Section \ref{boxes}. Therefore systems with constant DOS and total energy of the type (\ref{conf_en}) {\em do not} represent a realisation of the fractional exclusion statistics, at least not in any sense defined so far.

\subsection{2D Bose gas in MFA' approximation: an unexpected condensation}

To continue exercising with the MFA' approximation, let me now 
consider a 2D interacting Bose gas. I assume that the particle-particle 
interaction is repulsive, so $H_{\rm I}(N)\equiv\sigma^{-1}gN$, with $g>0$. 
Under canonical condition and in the MFA approximation, the gas is 
thermodynamically equivalent with the ideal 
Bose gas and its entropy and specific heat are given by the universal 
expressions (\ref{entropia}) and (\ref{cv}). On the other hand, in the 
MFA' model the energies of the 
single particle states are ``shifted'' according to Eq. (\ref{pseudo}), 
but the occupation of the single particle states is done 
in accordance with the ``original'' Bose statistics (see Eq. \ref{dim}). 
If I number the single particle (kinetic) energies and pseudoenergies as 
$\epsilon_0=0<\epsilon_1\le \ldots$, and $\eps_0<\eps_1\le \ldots$, 
respectively (I assume level 0 is nondegenerate), then the MFA' 
grandcanonical partition function is 
\begin{eqnarray}
\tZ(\beta,\beta\mu) &\equiv& \sum_{\{n_i\}} \rme^{-\beta\sigma^{-1}gh_0n_0^2
+\beta\mu n_0} 
\cdot \prod_{i\ge 1} \rme^{-\beta \eps_in_i+\beta\mu n_i}  \label{ZMFA'} \\
&\equiv& \sum_{n_0}\rme^{-\beta\sigma^{-1}gh_0n_0^2+\beta\mu n_0} 
\cdot \tZE^{\{n_i\}} \nonumber 
\end{eqnarray}
where 
$\tZE^{\{n_i\}} \equiv \sum_{\{n_i\}}\prod_{i\ge1}\rme^{-\beta \eps_in_i+\beta\mu n_i}$, while $n_i$ represents the population of the single 
particle state $i$.
The exponent $-\beta\sigma^{-1}gh_0n_0^2+\beta\mu n_0$ has a maximum for 
$n_0\equiv n_{0,{\rm max}}=\sigma\mu\cdot (2gh_0)^{-1}$. If $\mu<0$, 
$n_{0,{\rm max}} < 0$, and I expect no macroscopic population of any 
single particle state. In such a case, according to Eq. (\ref{dim}), 
$\trho(\eps)=\sigma n_{g}(\eps)$ and I should recover the 
previous results for the ideal Haldane system of parameter $g$ 
(see \cite{wu}). 
This will be shown to be true and for the Haldane gas of parameter $g>0$ 
(with the chemical potential $\mu_{\rm H}$) under canonical 
conditions, exists a strictly positive temperature 
$T_{\rm inv}$ (defined by Eq. \ref{mu}), so that 
$\mu_g(T<T_{\rm inv})>0$. 
Since for $T>T_{\rm inv}$, $\mu=\mu_{\rm H}$, I conclude that 
$\mu(T>T_{\rm inv})\to 0$ as $T\to T_{\rm inv}$.
Turning back to the MFA' model in grandcanonical ensemble, I 
investigate the situation $\mu\ge 0$. If $\mu>0$, the distribution 
$\rme^{-\beta\sigma^{-1}gh_0n_0^2+\beta\mu n_0}$ has a 
very sharp maximum centred at $n_0 = n_{0,{\rm max}}>0$ and with 
the second moment 
$\delta^2n_0=(\beta\mu)^{-1}n_{0,{\rm max}}$.
The relative fluctuation of $n_0$ is 
$\sqrt{\delta^2n_0}/n_{0,{\rm max}}=(\beta\mu)^{-1/2}n_{0,{\rm max}}^{-1/2}$, 
which is zero in the thermodynamic limit. 
Most often, such a sharp distribution is believed to fix the average value 
of the variable at the distribution maximum, but this is an example 
where this general procedure would lead to wrong results.

For a rigorous calculation, I denote by 
$\Z_g(\beta,\beta\mu)$ the grandcanonical partition function of the 
Haldane system of parameter $g$ and I assume for the beginning 
that $n_{i\ge 1}$ are all microscopic populations (in the end 
this will turn out to be true). Then %$\tZ(\beta,\beta\mu)$ becomes 
\begin{eqnarray}
\fl
\tZ(\beta,\beta\mu) &\equiv& \sum_{n_0}\tZ_{n_0}(\beta,\beta\mu)
=\sum_{n_0}\rme^{-\beta\sigma^{-1}gh_0n_0^2
+\beta\mu n_0} \cdot \Z_g[\beta,\beta(\mu-\eps_1)] \nonumber \\
\fl
&\approx& \sum_{n_0}\rme^{-\beta\sigma^{-1}gh_0n_0^2+\beta\mu n_0} 
\cdot \Z_g\left[\beta,\beta\left(\mu-\frac{gn_0}{\sigma}\right)\right] 
. \label{raw_part}
\end{eqnarray}
The most probable value of $n_0$ in the summation (\ref{raw_part}) 
is given by the equation $\partial \tZ_{n_0}/\partial n_0=0$. 
To find this value I define the function 
\begin{equation} \label{pop_0}
f_{h_0}(n_0,\mu)\equiv \tZ_{n_0}^{-1}\frac{\partial \tZ_{n_0}}{\partial n_0} =
-2h_0n_0\frac{\beta g}{\sigma}+\beta\mu - \langle N_{\rm ex}\rangle 
\frac{\beta g}{\sigma} ,
\end{equation}
where $\langle N_{\rm ex}\rangle\equiv\partial\log(\Z_g)/\partial(\beta\mu')$ 
is the average number of particles on the excited energy levels. 
The most probable value of $n_0$ is given by one of the zeros of $f$. 
For simplicity I shall use the notations $\xi\equiv\beta n_0g/\sigma$, 
$\zeta\equiv \beta\mu$, and $\mu'\equiv \mu-\sigma^{-1}gn_0$. 
Plugging Eq. (\ref{N}) into (\ref{pop_0}) I can rewrite $f$ in 
two equivalent ways:
\begin{eqnarray}
\fl
f_{h_0}(\xi,\zeta) &=& \log\left(\frac{y'_0}{1+y'_0}\right)+(1-2h_0)\xi  
\label{f1} \\
\fl
&=& (1-2h_0)\zeta + 2h_0\log(y'_0)-[2h_0+g(1-2h_0)]\log(1+y'_0) ,
\label{f2}
\end{eqnarray}
where $(1+y'_0)^{1-g}/y'_0=\rme^{-(\zeta-\xi)}$. 
Expression (\ref{f1}) implies that $f_{h_0}(0,\zeta)<0$ for any $h_0$ and 
$f_{h_0}(\xi,\zeta)<0$ for any $\xi$, if $h_0\ge 1/2$. 
Therefore the probability distribution of $n_0$ has a local maximum at 
$n_0=0$. If $h_0\ge 1/2$, then $n_0=0$ is the only maximum of $f_{h_0}$. 
Since 
$\tZ_{n_0}=\Z_g(\beta,\beta\mu)\cdot\exp\left(g^{-1}k_{\rm B}T\sigma\int_0^{\xi(n_0)}f_{h_0}(\xi',\zeta)\rmd\xi'\right)$ and $k_{\rm B}T\sigma$ 
is assumed to be a macroscopic quantity, it follows that 
$\tZ_{n_0}$ has an infinitely sharp maximum at $n_0=0$ which implies that 
the ground state is microscopically populated in spite of the 
very sharp maximum of $\rme^{-\beta\sigma^{-1}gh_0n_0^2+\beta\mu n_0}$, 
centred at $n_{0,{\rm max}}$.

To observe the behaviour of $f_{h_0}$ for 
large $\xi$ we have to look at the expression (\ref{f2}). For 
$h_0=0$, the function gets the simple form 
$f_0(\xi,\zeta) = \zeta -g\log(1+y'_0)$. Since $y'_0(\zeta-\xi)$ decreases 
monotonically from $y'_0(\zeta)$ to $y'_0(-\infty)=0$, as $\xi$ increases 
from $0$ to $\infty$, then $f_0(\xi,\zeta)$ is also a monotonic function of 
$\xi$, with $f_0(0,\zeta)<0$ and $f_0(\infty,\zeta)=\zeta>0$. 
The only zero of $f_0$ is at 
$y'_0=y'_{0,{\rm min}}\equiv\rme^{\zeta/g}-1$ and corresponds 
to a minimum of probability. As one can observe directly from Eq. 
(\ref{ZMFA'}), the maximum probability is $\infty$ and corresponds 
to $n_0=\infty$. Yet, as I mentioned above, the system has a metastable 
state for microscopic $n_0$, which corresponds to the equivalent (ideal)
Haldane distribution. 

The function $f_{h_0}(\xi,\zeta)$ is a continuous function in all the 
parameters and variables. From Eq. (\ref{f1}) follows that for 
fixed $\xi$ and $\zeta$, $f$ is a decreasing function of $h_0$ 
and from Eq. (\ref{f2}) we observe that for $\xi\gg\zeta$, 
$f_{h_0}(\xi,\zeta)\approx\zeta-2h_0\xi$. For small enough $h_0$, 
$f_{h_0}(\xi,\zeta)=0$ has a solution at 
$\xi_{\rm max}\approx\zeta/(2h_0)$ and 
$\exp(\zeta-\xi_{\rm max})=\exp\left\{-\left[(2h_0)^{-1}-1\right]\zeta\right\}\ll \exp(\zeta/g)-1$. In such a case, and since 
$f_{h_0>0}(\xi\to\infty,\zeta)\to-\infty$ and also 
$f_{h_0}(0,\zeta)<0$ I can conclude that $f$ has two zeros. The first 
corresponds to a local minimum of probability for $n_0$, while the second 
to a local maximum. Whether $n_0=0$ or 
$n_0=n_{0,{\rm max}}\equiv\xi_{\rm max}\sigma/(\beta g)\approx\sigma\mu\cdot (2gh_0)^{-1}$ 
has higher probability, it depends 
on the specific values of $h_0$ and $\zeta$. Nevertheless, continuity 
of $f$ and monotonicity in $h_0$ implies that exists a critical 
value of $h_0$, between 0 and 1/2 and which 
increases monotonically with $\zeta$, I call it $h_{0,{\rm cr}}(\beta\mu)$, 
so that for $h_0<h_{0,{\rm cr}}(\beta\mu)$ the ground state is 
macroscopically populated 
(the probability for $n_0=n_{0,{\rm max}}$ is highest), while 
for $h_0>h_{0,{\rm cr}}$ the ground state has microscopic occupation number
and is well described by the Haldane ideal gas of parameter $g$. 
If the ground state is macroscopically populated, I say that 
{\em the 2D MFA' Bose system is condensed}.

Now I prove that none of the excited energy 
levels are macroscopically occupied. Obviously, I start with $\eps_1$. 
This level can be macroscopically occupied {\em if and only if} 
$\mu'(n_{0,{\rm max}})\ge0$ 
and $h_0\le h_{0,{\rm cr}}(\beta\mu')$. The condition $\mu'\ge0$ 
implies $h_0\le h_{0,{\rm cr}}(\beta\mu)$ and 
$n_{0,{\rm max}}\le g^{-1}\sigma\mu$. But for 
$n_{0}=g^{-1}\sigma\mu$, $\xi=\zeta$, and 
$f_{h_0}(\zeta,\zeta)=\log[y_0(0)]-\log[1+y_0(0)]+(1-2h_0)\zeta$ -- 
where $y_0(0)>1$ is a fixed value which satisfy the equation 
$y_0(0)\cdot(1+y_0(0))^{g-1}=1$. 
Therefore $f_{h_0}(\zeta,\zeta)$ is a function linear in $\zeta$, 
which starts at $f_{h_0}(0,0)=\log[y_0(0)]-\log[1+y_0(0)]<0$ and, 
since $h_0\le h_{0,{\rm cr}}(\beta\mu)<1/2$, 
increases to infinity, as $\zeta$ increases. 
The continuity of $f_{h_0}(\xi,\zeta)$ completes then the proof 
that always $\zeta<\xi_{\rm max}$, which implies that 
$n_{0,{\rm max}}> g^{-1}\sigma\mu$. This proves the fact that 
$\eps_1$ and as a consequence any state of pseudoenergy 
$\eps_{i\ge 1}$ is microscopically populated. Also, the monotonic increase 
of $h_{0,{\rm cr}}$ with $\zeta$ implies that, if the 
ground state is microscopically populated, so are all the other states.

To finish the exercise I will show that for any $0\le h_0<1/2$, 
exists a temperature $T_{\rm c}<T_{\rm inv}$ at which the system 
condenses on the ground state. If the system is not condensed, it is 
described as a Haldane gas of parameter $g$. In this case 
the relative fluctuation of the total particle number $N$ 
vanishes in the thermodynamic limit. On the other hand, for a 
condensed system, 
\begin{eqnarray}
\tZ_{n_0} &=& \tZ_{n_{0,{\rm max}}}\cdot\exp\left(
\frac{k_{\rm B}T\sigma}{g}
\int_{\xi_{\rm max}}^{\xi(n_0)}f_{h_0}(\xi',\zeta)\rmd\xi'\right) 
\nonumber \\
&\approx& \tZ_{n_{0,{\rm max}}}\cdot\exp\left(
\frac{\beta g}{2\sigma}(n_0-n_{0,{\rm max}})^2
\left.\frac{\partial f_{h_0}(\xi,\zeta)}{\partial \xi}
\right|_{\xi=\xi_{\rm max}}\right) , \label{n0_fluct}
\end{eqnarray}
where I used the fact that $f_{h_0}(\xi_{\rm max},\zeta)=0$. 
From (\ref{n0_fluct}) it is easy to observe that the relative 
fluctuation is proportional to $n_{0,{\rm max}}^{-1/2}$, which 
vanishes in the thermodynamic limit. The relative total particle fluctuation,
which is obtained by adding together the contributions from the ground 
state, from the excited states (described as a Haldane gas), and the 
correlations between them, vanishes also. It is therefore safe to 
use the grandcanonical average values in canonical calculations 
even for this unusual toy system (fine-tuning due to the 
change of ensemble are relevant only for the finite size effects).
At $T=T_{\rm inv}$, $\mu=0$ and so is $\zeta$. In this case, from 
Eq. (\ref{f2}) we see that $f_{h_0}(\xi,0)<0$ for any $\xi$ and 
$h_0$, therefore $n_0$ is still microscopic, so the condensation 
temperature is lower. If the system does not condense, $\mu$ increases 
as the temperature is lowered, which implies an even faster increase 
of $\zeta$. If I assume that for a chosen $h_0<1/2$, $T_{\rm c}=0$ 
(the system does not condense), then I can choose $T$ so that 
$\zeta$ takes any value up to infinity. If $\zeta-\xi\gg 1$, 
then $y'_0\approx \rme^{(\zeta-\xi)/g}$ and 
$\log[(1+y'_0)/y'_0]\approx\rme^{-(\zeta-\xi)/g}\ll 1$. 
Now, for any $h_0<1/2$ and interval $[0,\xi_0]$, I can 
choose $\zeta\gg 1+\xi_0$, so that $f_{h_0}(\xi_0,\zeta)>0$ and 
$\int_0^{\xi_0}\log[(1+y'_0)/y'_0] < (1-2h_0)\xi_0^2/2$, 
which implies that 
$\tZ_{n_{0,{\rm max}}}>\tZ_{n_0(\xi_0)}>\tZ_{0}=\Z_g(\beta,\beta\mu)$. 
In this case the system is condensed, with $n_{0,{\rm max}}$ particles 
on the ground state, so the initial assumption ($T_{\rm c}=0$) was 
wrong. 

In conclusion, for any $h_0<1/2$, exists a temperature 
$T_{\rm c}\in(0,T_{\rm inv})$, 
at which the condensation occurs and bellow which the similarity 
between the MFA' Bose system and the ``usual'' ideal 
Haldane system \cite{wu} of parameter $g$ is lost. Moreover, the 
onset of the condensation removes also any thermodynamic 
equivalence between gases described in MFA and MFA' models. The exercise 
presented here is also interesting for the fact that it showed in a 
concrete example 
how an infinitely sharp (in the thermodynamic sense) 
probability distribution of particle on the ground state may 
be overwhelmed by the probability distribution of particles on the 
excited energy levels. Vice versa, it also shows that the equilibrium 
distribution of the Haldane gas, as it was deduced by 
Isakov \cite{isakov1,isakov3} and Wu \cite{wu}, and which corresponds 
to microscopic $n_0$, may not be the equilibrium distribution, 
in spite of the FES manifested in Eq. (\ref{dim}).

\end{document}